\documentclass[figures]{epl}
\usepackage{amsmath}
\usepackage{amssymb}

\title{Structural relaxation in a supercooled molecular liquid}
\author{S.-H. Chong \and F. Sciortino}
\institute{
Dipartimento di Fisica and
Istituto Nazionale per la Fisica della Materia, 
Center for Statistical Mechanics and Complexity,
Universit\`a di Roma ``La Sapienza'' -
Piazzale Aldo Moro 2, I-00185, Roma, Italy}


\begin{document}

\maketitle

\begin{abstract}

We perform molecular-dynamics simulations of a 
molecular system in supercooled states for different 
values of inertia parameters to provide evidence 
that the long-time dynamics depends only on the 
equilibrium structure. 
This observation is consistent with the prediction of 
the mode-coupling theory for the glass 
transition and with the hypothesis that the potential 
energy-landscape controls the slow dynamics. 
We also find that 
dynamical properties at intermediate wavenumber
depend on the spatial correlation of the molecule's 
geometrical center.

\end{abstract}

Dynamics in normal liquid states deals with orbits in phase space.
Binary collisions and vibrations,
for which the kinetic energy plays an important role, 
are elementary ingredients
building the motion on microscopic time and frequency scales.
In supercooled states, 
the system gets trapped in potential energy-landscape pockets for
long times.
This produces a separation of the long-time glassy dynamics ---
corresponding to the interbasin dynamics, i.e., to the 
motion from one pocket to the other --- 
from the microscopic dynamics.

In recent years the mode-coupling theory (MCT) for the
glass transition has been developed~\cite{Goetze91b}.
One of its essential predictions is the independence of 
the glass-transition singularity, as well as of
the long-time glassy dynamics, 
from the microscopic dynamics. 
This means that the glass-transition singularity and 
associated slow dynamics are completely
determined by the statistics of the system's
orbits in configuration space rather than in phase space~\cite{Franosch98}.
Such dynamics, determined solely by the equilibrium structure,
is referred to as the structural relaxation.
This MCT prediction agrees with the
hypothesis that the potential energy-landscape controls the slow 
dynamics~\cite{Stillinger95}. 
Indeed, 
an interesting characterization of the MCT critical temperature
$T_{c}$ in terms of geometrical properties of 
the landscape has been proposed~\cite{Angelani00,Broderix00}.

The issue of the independence of the long-time dynamics from underlying
microscopic dynamics has been addressed for a binary
mixture of Lennard-Jones particles~\cite{Gleim98}.
In Ref.~\cite{Gleim98}, molecular-dynamics (MD) simulations have been 
performed based on Newtonian as well as stochastic dynamics,
and the same long-time relaxation was observed. 
One of the main points of this Letter is to explore whether
such a property holds for molecular systems.
This is nontrivial because the dynamics in molecular
liquids involves rotational as well as translational motions,
and there exist more than two time scales characterizing 
microscopic dynamics. 
Since these microscopic motions depend on molecule's inertia 
parameters, 
we report in this Letter MD simulations for systems 
which differ in these parameters. 
Let us note that MCT for molecular systems --
both the molecular theory based on the
expansion in generalized spherical harmonics~\cite{tensor-theory} 
and the site-site theory~\cite{site-theory} --
predicts the inertia-parameter independence of the long-time
dynamics, and our work also serves as the first test of 
such a theoretical prediction. 

Another motivation of the present study concerns 
the unusual wavenumber dependence of characteristic time 
in molecular systems. 
A strong correlation 
in the wavenumber dependence 
is usually observed between the so-called
nonergodicity parameters (which specify the strength of the so-called
$\alpha$ relaxation),  
the $\alpha$-relaxation times, and the
static structure factor.
This is found, e.g., in hard-sphere system~\cite{Fuchs92b},
Lennard-Jones binary mixture~\cite{Kob95b}, 
water~\cite{Sciortino97} and silica~\cite{Horbach01,Sciortino01}, 
and is in agreement with the MCT prediction. 
The result found in a MD simulation study for a 
model of orthoterphenyl (OTP)~\cite{Rinaldi01}
is unusual in that such a correlation is violated for wavenumbers
about half the position $q_{\rm max}$
of the first sharp diffraction peak. 
It was suggested in Ref.~\cite{Rinaldi01} that the unusual feature 
is caused by the coupling of the rotational motion to the
center-of-mass motion. 
By changing inertia parameters,
it shall be examined whether it is the coupling to the
geometrical center or to the center of mass which matters. 

We consider a model for OTP 
designed by Lewis and Wahnstr\"om (LW)~\cite{Lewis94}
and a modified model which differs only in 
the mass distribution. 
The LW molecule is a rigid isosceles triangle.
Each of the three sites represents an entire phenyl ring of 
mass $m \approx 78\un{amu}$, 
and is described by a Lennard-Jones sphere. 
Extensive studies on the dynamics and thermodynamics of the original model
have recently been performed~\cite{Rinaldi01,Mossa02}.
In the modified model,  
the mass of the central site, $m_{c}$, and that
of the two adjacent sites, $m_{s}$, are chosen so that
$m_{c}/m_{s} = 16$ but with the total mass being unchanged,  
$m_{c} + 2 m_{s} = 3m$.
The center-of-mass position in the modified molecule
nearly coincides with
the position of the central site. 
The thermal velocity for the center-of-mass 
translation is the same for both models.
However, the thermal angular velocities are increased for the
modified model relative to the original one.
Thus, one expects a considerable speeding up of the reorientational
motion, and, via the rotation-translation coupling, a
considerable change in the complete microscopic dynamics. 
We have studied, performing MD simulations, 
both the original and modified systems
composed of $N = 343$ molecules at the density
$\rho = 1.083\un{g/cm^{3}}$, with the same procedure
described in Ref.~\cite{Mossa02}. 

Three normalized density correlators
shall be considered: 
\begin{equation}
\phi_{q}^{\rm X}(t) = F_{q}^{\rm X}(t) / F_{q}^{\rm X}(0) \quad
\mbox{for} \quad \mbox{X = N, Z, and Q}.
\end{equation}
Here, $F_{q}^{\rm X}(t)$ are defined by
$F_{q}^{\rm X}(t) = 
\langle \rho_{\vec q}^{\rm X}(t)^{*} \rho_{\vec q}^{\rm X}(0) \rangle / N$
in terms of the linear combinations 
\begin{equation}
\rho_{\vec q}^{\rm N} = 
(\rho_{\vec q}^{1} + \rho_{\vec q}^{2} + \rho_{\vec q}^{3}) / \sqrt{3}, \quad 
\rho_{\vec q}^{\rm Z} = 
(2\rho_{\vec q}^{1} - \rho_{\vec q}^{2} - \rho_{\vec q}^{3}) / \sqrt{6}, \quad
\rho_{\vec q}^{\rm Q} = 
(\rho_{\vec q}^{2} - \rho_{\vec q}^{3}) / \sqrt{2},
\end{equation}
of the site-density fluctuations
$\rho_{\vec q}^{a} =
\sum_{i=1}^{N} \exp(i {\vec q} \cdot {\vec r}_{i}^{\, a})$
($a=1$, 2, or 3)
in which ${\vec r}_{i}^{\, a}$ denotes the position of site $a$
in $i$th molecule.
(We use the convention that $a=1$ refers to the central site 
and $a=2,3$ to the two adjacent sites.)
Assuming equal scattering lengths, the correlator
$\phi_{q}^{\rm N}(t)$ is directly related
to the cross section as measured in the coherent neutron scattering. 
The functional forms of $\phi_{q}^{\rm Z}(t)$ and $\phi_{q}^{\rm Q}(t)$
have been chosen so that their
small-wavenumber limits reduce to the 1st-rank 
reorientational correlators~\cite{note-C1}. 
Dynamical features of $\phi_{q}^{\rm N}(t)$ for the original LW OTP
have already been discussed in Ref.~\cite{Rinaldi01}.

\begin{figure}
\onefigure[width=0.5\linewidth]{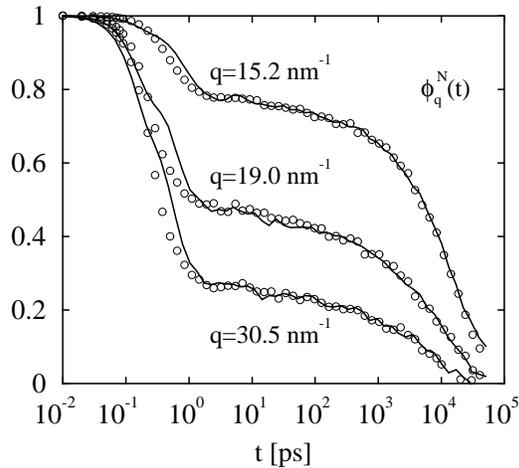}
\caption{
The correlators $\phi_{q}^{\rm N}(t)$ 
at $T = 260$ K for wavenumbers
$q = 15.2$, $19.0$, and $30.5\un{nm^{-1}}$.
Circles and solid lines respectively refer to the results for the
original and modified LW OTP, 
and the time scale of 
the latter is rescaled as $t/\hat{t}_{0}$ with $\hat{t}_{0} \approx 0.71$.} 
\label{fig:1}
\end{figure}

\begin{figure}
\onefigure[width=0.5\linewidth]{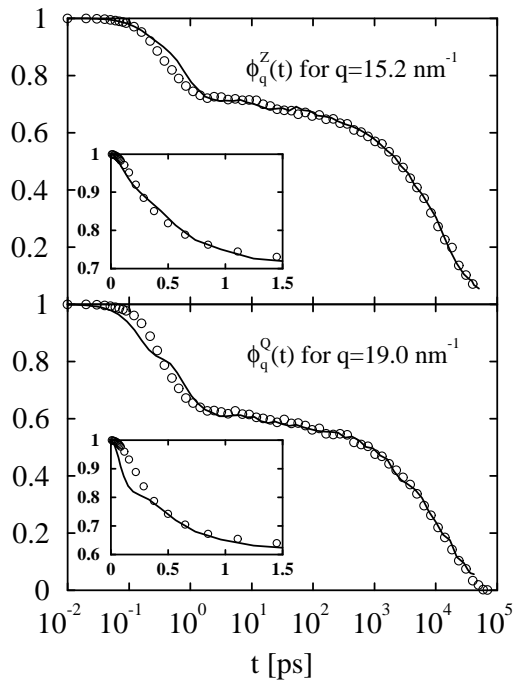}
\caption{
The correlators $\phi_{q}^{\rm Z}(t)$ for 
$q = 15.2\un{nm^{-1}}$
(upper panel)
and $\phi_{q}^{\rm Q}(t)$ for $q = 19.0\un{nm^{-1}}$ (lower panel)
at $T = 260\un{K}$. 
Circles and solid lines respectively refer to the results for the
original and modified LW OTP, 
and the time scale of the
the latter is rescaled by $\hat{t}_{0} \approx 0.71$. 
The insets show the short-time dynamics,
{\em without rescaling the time scale for the modified system}, 
on the linear time axis.}
\label{fig:2}
\end{figure}

We show in Fig.~\ref{fig:1} 
the correlators $\phi_{q}^{\rm N}(t)$
at $T = 260\un{K}$ for wavenumbers
$q = 15.2$, $19.0$ and $30.5\un{nm^{-1}}$. 
(We notice that the first peak in the static structure factor
$S_{q}^{\rm N} = F_{q}^{\rm N}(0)$ 
is located at $q_{\rm max} = 14.6\un{nm^{-1}}$.)
Circles and solid lines denote the results for
the original and modified systems, respectively, and
the time scale of the latter
is rescaled by a factor $\hat{t}_{0} \approx 0.71$, whose meaning will be
discussed below. 
Except for $t \lesssim 1\un{ps}$, curves for the two systems coincide
within the statistical errors, 
i.e., the long-time dynamics is, 
up to the scale $\hat{t}_{0}$, 
independent of inertia parameters. 
Figure~\ref{fig:2} shows corresponding results for $\phi_{q}^{\rm Z}(t)$
and $\phi_{q}^{\rm Q}(t)$. 
While the short-time dynamics exhibits inertia 
dependence (see insets of Fig.~\ref{fig:2}), 
the long-time dynamics is identical up to the same scale $\hat{t}_{0}$.

The findings in Figs.~\ref{fig:1} and \ref{fig:2}
verify the following formulation of the
glassy dynamics, as suggested by MCT~\cite{Goetze91b,Franosch98}:
the microscopic dynamics determines a time scale $t_{0}$
so that for $t \gg t_{0}$ any correlator $\phi_{x}(t)$
can be written as
\begin{equation}
\phi_{x}(t) = F_{x}(t/t_{0}).
\label{eq:structural}
\end{equation}
Here, the master function $F_{x}(\tilde{t})$ is independent of the 
microscopic dynamics,
and the scale $t_{0}$ is common to all the correlators. 
The slow dynamics, as described by $F_{x}(\tilde{t})$,
is solely determined by the potential energy $V$ 
through the Boltzmann factor $e^{- V/k_{B}T}$
(with $k_{B}$ denoting the Boltzmann constant),  
i.e., by the statistics of orbits in configuration space.
The microscopic dynamics, which depends on inertia parameters,
merely sets the scale $t_{0}$ for the exploration of 
the potential energy-landscape. 
Thus, the factor $\hat{t}_{0}$ used in 
Figs.~\ref{fig:1} and \ref{fig:2} can be understood
as the ratio $\hat{t}_{0} = t_{0}^{\rm II}/t_{0}^{\rm I}$
of the $t_{0}$'s for the original ($t_{0}^{\rm I}$)
and modified ($t_{0}^{\rm II}$) systems.
The region where circles and solid lines 
coincide is the structural-relaxation
part as described by $F_{x}(\tilde{t})$. 
The speeding up of the rotational motion for the modified system,
demonstrated in the insets of Fig.~\ref{fig:2},
means that the potential energy-landscape can be explored faster.
This explains why $\hat{t}_{0} < 1$. 

\begin{figure}
\onefigure[width=0.5\linewidth]{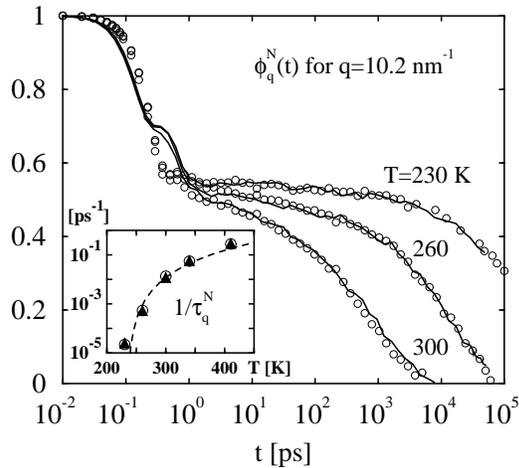}
\caption{
The correlators $\phi_{q}^{\rm N}(t)$ for
$q = 10.2\un{nm^{-1}}$
at three temperatures $T = 230$, $260$, and $300\un{K}$.
Circles and solid lines respectively refer to the results for the
original and modified LW OTP, 
and the time scale of the
the latter is rescaled by $\hat{t}_{0} \approx 0.71$.
The inset exhibits the inverse of the $\alpha$-relaxation times
$1/\tau_{q}^{\rm N}$ as a function of temperature:
$1/\tau_{q}^{\rm N}$ for the original LW OTP are marked with
open circles, while those for the modified system 
(rescaled by $\hat{t}_{0} \approx 0.71$) with filled triangles. 
Dashed line denotes the fit based on the
prediction of MCT,
$1/\tau_{q}^{\rm N} \propto (T - T_{c})^{\gamma}$,
with $T_{c} \approx 234\un{K}$ and $\gamma \approx 2.76$.}
\label{fig:3}
\end{figure}

Figure~\ref{fig:3} provides another test of 
Eq.~(\ref{eq:structural}), where we plot
$\phi_{q}^{\rm N}(t)$ for $q = 10.2\un{nm^{-1}}$
at three temperatures close to the MCT critical temperature
$T_{c} \approx 234\un{K}$~\cite{Mossa02}. 
The same $\hat{t}_{0} \approx 0.71$ has been used 
to rescale the correlators for the modified system.
Since the microscopic dynamics can also depend on $T$, 
a smooth variation of $\hat{t}_{0}$ with $T$ is expected:
for example, 
we checked that, by slightly modifying the value of $\hat{t}_{0}$,
an even better agreement in the long-time dynamics is obtained for
$T = 300\un{K}$. 
However, the temperature dependence of $\hat{t}_{0}$ is rather
small close to $T_{c}$ as can be seen from Fig.~\ref{fig:3}.
We note that, within MCT, $\hat{t}_{0}$ should be evaluated at 
$T = T_{c}$~\cite{Goetze91b,Franosch98}.

The dynamics in supercooled states exhibits two-step
relaxation as shown in Figs.~\ref{fig:1}--\ref{fig:3}:
the relaxation toward the plateau, followed by the final relaxation
from the plateau to zero (the $\alpha$ relaxation).
The structural-relaxation region, where circles and solid
lines coincide in Figs.~\ref{fig:1}--\ref{fig:3}, is precisely the one
for which MCT is derived and its applicability should be tested.
We notice from Figs.~\ref{fig:1}--\ref{fig:3}
that the approach toward the plateau,
for which MCT predicts an asymptotic power-law decay
$\sim t^{-a}$ ($0 < a < 0.5$)~\cite{Goetze91b}, 
is almost completely masked by the microscopic dynamics. 
This implies that the most faithful tests of MCT should be performed
just near the plateau and in the $\alpha$-relaxation regime.
A similar conclusion was drawn in Ref.~\cite{Sciortino99}
in a different context. 

The inset of Fig.~\ref{fig:3} shows 
$1/\tau_{q}^{\rm N}$ of the $\alpha$-relaxation times
for the original and modified systems.
In this work, $\tau_{q}^{\rm N}$ are defined by
$\phi_{q}^{\rm N}(\tau_{q}^{\rm N}) = 1/e$. 
MCT predicts an asymptotic formula, 
$1/\tau_{q}^{\rm N} \propto (T - T_{c})^{\gamma}$, 
close to $T_{c}$~\cite{Goetze91b}. 
The inset shows that
the asymptotic formula describes well the result
with $T_{c} \approx 234\un{K}$ and $\gamma \approx 2.76$
as estimated in Ref.~\cite{Mossa02}.
The exponent $\gamma$ is uniquely related to the so-called 
exponent parameter $\lambda$~\cite{Goetze91b}.  
The result indicates that, also for molecular systems, 
$T_{c}$ and $\lambda$
are independent of inertia parameters, in agreement with the 
prediction of MCT. 
Notice that this conclusion does not depend on the quality of the
fit (i.e., the choice of the values for $T_{c}$ and $\gamma$)
since, after the rescaling by $\hat{t}_{0}$, 
the correlators $\phi_{q}^{\rm N}(t)$ and hence $1/\tau_{q}^{\rm N}$
for the two systems coincide within the statistical errors. 

Equation~(\ref{eq:structural}) is based on the ideal MCT,
whose validity is restricted to $T > T_{c}$. 
More precisely, Eq.~(\ref{eq:structural}) is valid also below
$T_{c}$, but according to MCT, 
$F_{x}(\tilde{t})$ for $T \le T_{c}$ would describe the arrest 
at the plateau~\cite{Goetze91b}, 
and does not properly account for the $\alpha$
relaxation in the simulation results, 
e.g., the $T = 230\un{K}$ result shown in Fig.~\ref{fig:3}.
The ideal MCT predicts the inertia-parameter independence of 
the plateau height,
and this can be confirmed from the figure.
However, the $T = 230\un{K}$ result also shows that, 
even for $T < T_{c}$, 
the $\alpha$ relaxation seems to be independent
of inertia parameters. 
This implies the possibility of extending 
the formulation (\ref{eq:structural}) to $T < T_{c}$
including the whole $\alpha$-relaxation regime.
Such a finding might help in developing theories
for $T < T_{c}$. 
Let us note that 
this is also consistent with the
energy-landscape description whose applicability is not 
restricted to $T > T_{c}$. 

\begin{figure}
\onefigure[width=0.5\linewidth]{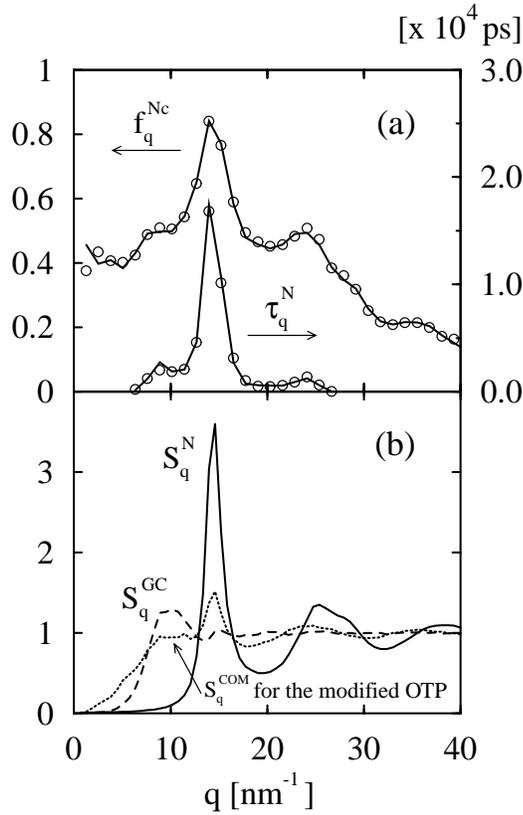}
\caption{
(a) 
The critical nonergodicity parameters $f_{q}^{{\rm N} c}$
and the $\alpha$-relaxation times $\tau_{q}^{\rm N}$ for the
correlators $\phi_{q}^{\rm N}(t)$. 
The latter are calculated at $T = 260\un{K}$. 
Circles and solid lines respectively refer to the results for the
original and modified LW OTP, 
and $\tau_{q}^{\rm N}$ for 
the latter are rescaled by $\hat{t}_{0} \approx 0.71$.
(b)
The static structure factors at $T = 260\un{K}$. 
Solid and dashed lines respectively denote the site-site
static structure factor $S_{q}^{\rm N}$
and the geometrical-center
static structure factor $S_{q}^{\rm GC}$,
which are the same for both the original and modified LW OTP.
For the original system, 
$S_{q}^{\rm GC}$ coincides with the center-of-mass static structure
factor $S_{q}^{\rm COM}$.
$S_{q}^{\rm COM}$ for the modified system is denoted as dotted line.}
\label{fig:4}
\end{figure}

Next we turn our attention to the wavenumber dependence
of the $\alpha$-relaxation time for wavenumbers
around $q_{\rm max}/2$. 
Figure~\ref{fig:4}(a) exhibits the
critical nonergodicity parameters $f_{q}^{{\rm N}c}$ and 
the $\alpha$-relaxation times $\tau_{q}^{\rm N}$
of the correlators $\phi_{q}^{\rm N}(t)$
for the original and modified systems. 
Here, $f_{q}^{{\rm N}c}$ are the plateau height 
in the two-step relaxation, and are
obtained from the
fit based on the von-Schweidler law with its
first correction as in Ref.~\cite{Rinaldi01}.
Figure~\ref{fig:4}(a) shows that
these quantities for
the two systems are the same within the statistical errors.
One also sees a clear correlation in the wavenumber
dependence of $f_{q}^{{\rm N}c}$ and $\tau_{q}^{\rm N}$:
there exist corresponding peaks and minima
in these quantities.

The site-site static structure factor 
$S_{q}^{\rm N} = F_{q}^{\rm N}(0)$ associated with 
$\phi_{q}^{\rm N}(t)$ is shown in Fig.~\ref{fig:4}(b). 
$S_{q}^{\rm N}$ is the same for both the original and modified
systems since it depends only on the Boltzmann factor.
For $q \gtrsim 15\un{nm^{-1}}$, 
the $q$-dependence of $f_{q}^{{\rm N}c}$ and $\tau_{q}^{\rm N}$ 
correlates also with that of $S_{q}^{\rm N}$. 
For $q < 15\un{nm^{-1}}$, on the other hand,
such a correlation is not observed, and 
there is an unusual peak at 
$q \approx 9\un{nm^{-1}}$
in $f_{q}^{{\rm N}c}$ and $\tau_{q}^{\rm N}$ which 
does not exist in $S_{q}^{\rm N}$. 
This is in contrast to results found, e.g., 
in hard-sphere system~\cite{Fuchs92b}, 
Lennard-Jones binary mixture~\cite{Kob95b},
water~\cite{Sciortino97}, and silica~\cite{Horbach01,Sciortino01}, 
for which such a correlation 
holds for the whole wavenumber regime. 

In Ref.~\cite{Rinaldi01}, it was suggested that the unusual feature
around $q \approx 9\un{nm^{-1}}$ can be interpreted
as being due to the coupling to the center-of-mass (COM)
dynamics.
This follows from the fact that 
the COM static structure factor,  
$S_{q}^{\rm COM} =
(1/N) \sum_{i,j}
\langle 
e^{- i {\vec q} \cdot ({\vec r}_{i}^{\, \rm COM} - {\vec r}_{j}^{\, \rm COM})}
\rangle$
with ${\vec r}_{i}^{\, \rm COM}$ denoting the
COM position, 
has a peak around $q \approx 9\un{nm^{-1}}$ 
as shown in Fig.~\ref{fig:4}(b). 
For the original model,
the COM position is identical to the
geometrical center (GC) defined by
${\vec r}_{i}^{\, \rm GC} = (1/3) \sum_{a=1}^{3} {\vec r}_{i}^{\, a}$. 
The latter is a structural property, i.e., independent of
inertia parameters, 
and the corresponding structure factor $S_{q}^{\rm GC}$,
defined in terms of the GC positions, 
is the same for the two systems
in contrast to $S_{q}^{\rm COM}$ ({\em cf}. Fig.~\ref{fig:4}(b)).
Since Fig.~\ref{fig:4}(a) shows that the peaks in 
$f_{q}^{{\rm N}c}$ and $\tau_{q}^{\rm N}$ are also a structural property,
we find --- from the correlation between 
$f_{q}^{{\rm N}c}$, $\tau_{q}^{\rm N}$, and $S_{q}^{\rm GC}$ ---
that it is more appropriate to interpret
the unusual peak around $q \approx 9\un{nm^{-1}}$ as being
caused by the coupling to the GC dynamics. 
Thus, the unusual peak reflects a property of the
potential energy-landscape, a subtle interplay of the
translational and rotational dimensions in
configuration space.
We notice that such a peak in the nonergodicity parameters
around $q \approx 9\un{nm^{-1}}$ can be found in the 
coherent neutron-scattering result for real OTP molecules,
where no clear peak shows up in the static structure 
factor~\cite{Toelle97}.

The position $q \approx 9\un{nm^{-1}}$ where the unusual peak
we discussed occurs is compatible with the inverse of the van der
Waals radius $r_{\rm W} = 0.37\un{nm}$ for OTP molecule~\cite{Bondi64},
i.e., it is connected to the overall size of the molecule. 
It is interesting to note that a similar unusual peak was found in a model for 
polymer around wavenumbers close to the inverse radius of 
gyration~\cite{Aichele01-all}. 
In Ref.~\cite{Aichele01-all}, the unusual peak is interpreted
in terms of polymer specific properties.
Such an interpretation might not be suitable, 
if the unusual peak -- found in OTP and polymer --
reflects the same physics. 
Thus, it will be a theoretical challenge to reproduce and explain 
such an unusual feature observed in molecular and polymer systems.

\begin{acknowledgments}

We thank W.~G\"otze for valuable comments. 
We acknowledge support from MIUR COFIN and FIRB
and from INFM PRA-GENFDT. 

\end{acknowledgments}


\begin{thebibliography}{10}
\bibitem{note-C1}
One can show that 
\protect$\lim_{q \to 0} \phi_{q}^{\rm Z}(t) \propto
\sum_{i,j} \langle {\vec e}_{i}(t) \cdot {\vec e}_{j}(0) \rangle / N$
describes the collective reorientation of the molecule's symmetry axis
\protect${\vec e}_{i}$, whereas 
\protect$\lim_{q \to 0} \phi_{q}^{\rm Q}(t)
\propto \langle {\vec f}_{i}(t) \cdot {\vec f}_{i}(0) \rangle$ refers to the
reorientation of the vector 
\protect${\vec f}_{i}$ connecting two end sites
belonging to the same molecule. 
Due to the 
\protect$C_{2v}$ symmetry of the model, 
\protect$\phi_{q}^{\rm Q}(t)$ is identical to its self part.
\end{thebibliography}
\end{document}